# The Shutdown Problem: How Does a Blockchain System End?

Mark Stuart Day
MIT / Jefelex Systems LLC
mday@alum.mit.edu

**Abstract**

We define and examine the *shutdown problem* for blockchain systems: how to gracefully end the system's operation at the end of its useful life. A particular focus is those blockchain systems that hold archival data of long-lived interest. We outline what it means to achieve a successful shutdown, and compare those criteria to likely end-of-life conditions in a generic blockchain system. We conclude that the decentralized nature of blockchain systems makes shutdown difficult, particularly if the system uses an unstable consensus like the Nakamoto consensus of Bitcoin. Accordingly, we recommend against using blockchain with unstable consensus for any data whose value is likely to persist beyond the life of the blockchain system. For any such systems that are already in operation, we recommend considering a hard fork to implement stable consensus. Such consideration needs to happen well in advance of the system's end of life.

## Introduction

Descriptions of blockchain systems often assume a roughly steady state in which there is an adequate supply of honest nodes. However, wise system design recognizes that at some point in the future the system will become obsolete and must be replaced or shut down. When we consider the overall system life cycle, there are intervals at the start of system operation and at the end of system operation when the steady-state conditions clearly do not apply.

These transitions are of no great importance for toy blockchain systems, or those dealing with highly transient data. However, a number of proposed and operational blockchain systems involve data that is likely to be of some importance over a long period of time. For example, MIT offers digital diplomas [DT17, M17]. One claimed advantage is that the credential will still be valid even if MIT itself no longer exists as an institution. Likewise, there are local governments using blockchain for public records such as birth and marriage certificates [AP19].

It is easy to assume that these kinds of records (birth certificates, marriage certificates, diplomas) are stable and reliable even when their storage system is reaching the end of its life. Correspondingly, it is a source of concern if we learn that such records may be subject to loss, corruption, or forgery.

Even if a particular data record itself is not of obvious long-term value, it may be important to preserve as part of an aggregate. For example, a small financial transaction two years ago may still be important for auditing or investigation.

For all of these blockchain systems with long-term data, we consider what it means for the blockchain system to end. In particular, we look at how we can distinguish between desirable and undesirable endings, and what options are available to mitigate the problems identified.

We first describe an informal notion of a good ending for a general information system (ignoring blockchain issues). Then we describe an informal model of a generic blockchain system. Even this generic blockchain system has aspects that make it hard or impossible to achieve a good ending. We consider some possible ways to adjust the generic blockchain system to support good endings. Finally, we discuss the implications of this analysis.

**What is a good ending?**

Consider some generic system (not necessarily a blockchain) storing long-term data. If such a system is to be replaced or decommissioned, we can identify some features that we expect to be true at the end of its life:

1. The stored information should be in a stable form. Since the system is not continuing to operate, there should not be additional changes to that information.
2. There should not be an ongoing commitment to operating or supporting the old system. However it is accomplished, there should be a sharp reduction in the cost of maintaining the stored information.

We can summarize these points by saying that a good ending for the system renders the stored information into a form that is *stable* and *cheap*. If the information is unstable, and/or if there is a substantial ongoing expense, we would not consider that to be a good ending.

(We note that this may be only a partial description of the requirements for a good ending: there may well be additional requirements that sensible people would demand. However, it seems clear that any ending that does not meet these requirements is not a good ending.)

For any information system, finding a good ending is the *shutdown problem*. Equivalently, we can say that solving the shutdown problem requires finding a good ending.

This description of the shutdown problem is applicable to any information system, and has no necessary relationship to blockchains. In the next section we describe blockchains, and in the following section we consider the shutdown problem for blockchains.

**System model**

We describe a generic blockchain system as follows:

1. A *community* of *nodes* jointly constructs a shared sequence.
2. The elements of the sequence are data *blocks* that are chained together via digital signatures.
3. Each node maintains its own local version of the sequence.
4. The local versions at two different nodes may be different.
5. The correct state of the sequence – and in particular, the content of a new block at the end – is determined via some form of consensus executed among the nodes.
6. Some nodes are honest, while others are dishonest. (Sometimes for simplicity we will assume that honest nodes are always honest; but in general, our model allows each node to choose dishonest behavior when it is advantageous.)
7. Correct operation of the system depends on having "enough" honest nodes, as determined by the specific consensus mechanism.
8. At least some of the information stored in blocks may be of interest to nodes other than the one that added the block, at a point in time later than the block's addition.

For the concerns we examine here, the specifics of signing don't matter. Likewise, more elaborate topologies than a single chain (side chains, branching) don't matter as long as there is some part of the system's operation that acts as a literal blockchain: that is, a chain of blocks.

This generic system model is broad enough to include both open and permissioned blockchains. In an open blockchain, any node at all can participate; in a permissioned blockchain there is some kind of gatekeeper determining which nodes can participate.

(A gatekeeper may make it harder for dishonest nodes to participate, but it is unlikely that any real gatekeeper can guarantee that only honest nodes participate. As we will see, part of the problem of ending a blockchain is that it may take only a single dishonest node to cause problems.)

The system model is broad enough to include a variety of different consensus mechanisms. In particular, we consider both stable and unstable consensus mechanisms.

With a stable consensus mechanism, the addition of a block is never subsequently revised; the chain may be extended but the existing chain is never revised at any node. With an unstable consensus mechanism, any block is subject to revision; the last block is relatively easy to revise, the previous block is much harder to revise, and so on with increasing difficulty to the beginning of the chain. The so-called Nakamoto consensus implemented in Bitcoin [N08] is an important example of an unstable consensus mechanism.

When the system includes a sufficient fraction of honest nodes, the overall system works correctly and we may refer to it as being in its "honest mode." When the system does not include a sufficient fraction of honest nodes, the overall system is open to corruption and failures, and we may refer to it as being in its "dishonest mode." For simplicity, we assume that there is only one correct/honest mode of operating and one incorrect/dishonest mode.

(For our purposes, it's sufficient to establish a single simple example of the shutdown problem. Although there might be some interesting aspects in scenarios with competing dishonest nodes, none of those scenarios will serve to negate the problems we identify.)

Given this system model, there are two key issues:

1. The "old" part of the ledger is not easily modified, while the "young" tail of the ledger is (relatively) easily modified.
2. Honest operation depends on having an adequate supply of honest nodes.

We next define some terminology to help describe whether there is an adequate supply of honest nodes.

**Smooth vs. lumpy; thick vs. thin**

We consider a *universe* of potential nodes for a blockchain system, and a *community* of those nodes that actually comprise the blockchain system at any point in time. Every node we consider is always part of the universe, but may or may not be part of the community.

We can characterize the system's operation as smooth or lumpy. When the system is *smooth*, it does not matter if we randomly add a new node to the community or remove an existing node from the community. Despite the change, the system continues to operate in its same mode (honest or dishonest). In contrast, when the system is *lumpy* there is at least one node in the universe whose addition or removal changes the mode of the system from honest to dishonest (or vice-versa).

Sometimes a lumpy system is caused by a node that has an unusual weight (by whatever measure is used in the consensus scheme). However, systems necessarily become lumpier as the size of community shrinks, even if the weight is divided evenly among all nodes. Accordingly, we can characterize a community of nodes in a blockchain system as *thin* (lumpy, or at risk of lumpiness) or *thick* (smooth and with no near-term risk of lumpiness).

We note that this thick/thin analysis becomes more complex if nodes are allowed to become dishonest. In particular, when there are relatively few nodes, an honest node may have a larger incentive to behave dishonestly.

**Thick and thin shutdowns**

It is plausible to have a shutdown problem with either a thick or thin community. A thick shutdown means that the system is moving to an ending even though there are still plenty of active nodes. This kind of shutdown can occur when there is a deadline imposed by some outside entity, perhaps due to regulatory changes. As we will see, a thick shutdown is easier.

A thin shutdown means that the system is moving to an ending partly or entirely because of the dearth of honest nodes. In those circumstances, we have to consider that the system may switch to dishonest operation at any moment, or indeed may have already entered that mode. A thin shutdown is harder, but is also probably the more likely situation.

**How does a blockchain system end?**

Let's consider a simple example of a good ending and see where it can go wrong. We make the following assumptions:
- consensus is stable
- the system is operating honestly
- the community is thick
- there is a "final block" that is readily understood as signifying the end of the chain.

As long as the system continues honest operation long enough to add the final block at the end of the chain, the system reaches a good ending. (To recap those criteria: the information stored is stable, there is no need for ongoing engagement, and there is no difficulty with establishing authenticity of the information).

Now let us take up each of the assumptions in turn, and note the associated problems if the given assumption does not hold.

*Unstable consensus*: The shutdown problem is notably harder with a consensus mechanism that is unstable, so that an apparent decision may be later reversed. One important example of an unstable consensus mechanism is the Nakamoto consensus underpinning Bitcoin. In a blockchain system using Nakamoto consensus (or something similar) the status of the last block in the chain is in some sense contingent. There is no explicit declaration that the last block is valid, although there can be an explicit declaration that some particular block is invalid. Instead, there is a kind of superposition of states: the last block of the chain *might* be the one we know about; or it *might* be some other one we don't yet know about, or *perhaps* one that we know about but have chosen to disregard in favor of the one we are treating as the last block.

This slightly weird fuzziness of the last block actually extends all the way to the first block, albeit with lower likelihoods as we move earlier in the chain. The degree of fuzziness depends on details of the consensus mechanism. Because the "truth" of the blockchain is determined by the community of nodes, it is possible for changes in that community to produce changes in the chain.

If an honest community reaches some kind of final statement in a blockchain with stable consensus, they can collectively leave the blockchain alone. They don't care if dishonest players append garbage after the final block. But if the consensus is unstable, they can never rely on that blockchain as an archive unless they remain engaged in it, and engaged in a way that outweighs any potentially competing dishonest nodes.

It is safe for honest nodes to abandon a blockchain that contains no information of any continuing interest. However, if that blockchain contains any information that might need to be preserved, it can **never** be abandoned by the honest nodes. Leaving the blockchain alone allows it to be rewritten by dishonest nodes.

We consider possible mitigations of these problems in a later section.

*Dishonest operation*: If the system is already operating dishonestly, or if the system starts to operate dishonestly before a final block is stable on the chain, then the final block may not be written or the chain may be otherwise corrupted.

*Thin community*: Even if the system is operating honestly, a single failure by an honest node may be enough to tip it into dishonest operation. Likewise, a single new dishonest node may be enough to change the system. Notice that in some cases an honest node may choose to become a dishonest node. As we've already noted, we expect that the behavior of the system will be lumpier when there are fewer nodes – independent of the consensus mechanism.

Although the usual concern with dishonest nodes focuses on weight within the consensus mechanism, it's worth noting that dishonest nodes could also flip the system into dishonest operation by subverting communications. With a thin community and multiple dishonest nodes, it's possible that an unsuspecting honest node is communicating entirely with dishonest nodes that could suppress or forge messages.

*No final block*: A typical blockchain system lacks any mechanism to add a "final block." Indeed, it is not clear that such a mechanism is a good idea other than when one is trying to solve a shutdown problem. It would seem to be a risk for possible mischief.

**Mitigation strategies**

In this section, we consider possible approaches to mitigate the problems identified in the previous section.

*Mitigating unstable consensus:* In a system using unstable consensus, the clearest solution would be to somehow substitute stable consensus. Unfortunately, such a replacement is difficult. First is the general problem of any hard fork: changing the system in an incompatible way necessarily means splitting the existing community, and only those nodes agreeing to the new consensus would be part of the new system. Second is the specific concern that the adoption of stable consensus would necessarily be determined by the existing unstable consensus. Particularly with a thin community, it might well be hard or impossible to get the desired result.

It may be hard to avoid a situation in which a new node, playing by the original rules, finds the "old" system. If we were concerned with continuing operation, we would also be concerned with the possibility that our hard-fork change was adopted by only a minority of the nodes. But here our specific goal (to shut down the system) is actually advantageous: it may be completely reasonable to have a split in the community between a stable, finished, archived system and the unstable, unfinished, and untrustworthy "live" system.

Now let's consider what is possible if for some reason we can't bring in stable consensus, and are stuck with an unstable consensus. Is there an alternative to permanent engagement with the system? There could be some lower-commitment way in which the blockchain is preserved – one that uses physical irreversibility, for example. We could envision an honest community that combines blockchain with archival technologies, so that some part of the blockchain could never be rewritten. The challenge is to decide when something is "solid enough" to become irreversible. Although there may be a workable solution, our informal exploration of this topic suggests that any fixed threshold could be exploited by dishonest players to invoke the permanent archiving incorrectly (and in an unfixable way). At a minimum, we complicate analysis of the system's current state and overall correctness: we introduce the possibility of a chain that somehow cannot be replaced, even though it is not part of the longest chain.

*Mitigating thin community:* The lumpiness caused by too few nodes may be avoidable at the start of the system by pursuing strategies to rapidly increase the number of nodes – for example, "air drop" approaches that effectively give away tokens or otherwise provide incentives for joining.

If we think of a plot of "number of nodes over time" for the entire life of the system, we know there is some initial ramp from zero and a final ramp to zero. We know how to steepen the slope of the initial ramp so as to move rapidly into smooth operation.

Unfortunately, we don't know how to steepen the slope of the final ramp. And although we can discuss the abstract concept of smooth vs. lumpy behavior, nodes in an actual community do not have any accurate and timely way to judge their blockchain's situation.

It doesn't seem that it makes any difference whether the nodes are balanced and weighed by effort or by stake – there is still a problem that as the number of nodes shrinks and the community becomes thinner, the system is vulnerable to dishonest behavior that does not occur when the community is thick.

(We could think of this by analogy with classical vs. quantum behaviors, and observe that classical behaviors depend on having enough atoms or particles so that statistical issues dominate. When the number of atoms or particles is small, then the statistics don't dominate and instead random individual choices or events matter, and weird quantum behaviors are observable.)

This appears to be a general observation about blockchain systems. It's well-known that there must be a prevalence of honest nodes for the system to work (honestly). But it seems less well-known that there must be a certain quantity of nodes (stakes, work) to avoid situations in which a single random node's behavior potentially tips the overall balance.

*Mitigating final block:* In a system without any clear marker for ending a chain, we could introduce such a marker as a hard fork. The discussion of difficulties resembles the one we presented above for mitigating unstable consensus.

**Archiving the blockchain?**

Part of blockchain's promise is to be a collective ledger. Unfortunately, when we consider archiving the ledger the situation becomes rather dubious. There is no straightforward way for today's blockchains to become low-overhead but stable archives.

First, we observe that conventional solutions based on a single archive are likely to be unattractive for information that is being stored on a blockchain. After all, if the designer of the system were willing to entrust the information to a single party, they could have built the information store in a much more straightforward fashion – no blockchain at all, just some kind of a database mediated by the trusted single party.

For some particular situation it may be acceptable to solve the shutdown problem in such a not-very-blockchain-like way; but in general, an archiving solution cannot depend on private copies or central authority. So we expect that part of ending a blockchain would include designating multiple archivists. Presumably this designation would itself be a record on the blockchain.

What does such an archivist do? One possibility is to simply snapshot the state of the blockchain, take a hash of the snapshot, and publish the snapshot's hash to the New York Times [H19]. As with any blockchain system, there would be some doubt about the validity of the last few blocks, but the bulk of the archive would be preserved. Between the internal consistency mechanisms of the blockchain and the published snapshot, we would be confident that the data had not been corrupted.

However, even if one is using one of these designated archives, there are natural concerns about provenance (how do we know that the data is connected to what was originally on the blockchain?) and authenticity (how do we know that the data is unaltered from what was originally on the blockchain?). If we genuinely don't trust any of the archivists to get it right, we potentially need to compare multiple copies of the archived data; however, there is no guarantee that those copies will be resolvable to a single version. We can check that the data within a single archive is consistent – we can reject it if it doesn't meet the correctness requirements of the particular blockchain. But we can't otherwise distinguish between two offered versions of a particular blockchain.  A live blockchain has some handy properties that are no longer available when we are looking at recorded versions of a blockchain.

There is an additional problem of redirecting from the live chain to the archive chain(s). It's not clear how you can do that reliably, given the other problems that these systems have.

Fortunately, we know that it is expensive to forge a blockchain, and the cost increases for older blocks. In many – perhaps most – cases, we do not need to be concerned about an attacker constructing an entirely forged chain. However, any such reassurance depends heavily on the value of the stored data, and the consequent potential reward from changing it. There are sometimes surprising results from carefully examining the economic incentives for attacks [A19].

**Analysis and discussion**

*No exit?*

In most cases, for most blockchain systems, the shutdown problem is hard or impossible. There is no straightforward way to declare that a history is finished and archived. If you do not care enough about the integrity of the data to keep working at it (and potentially recruiting others to assist you) then your history will potentially be corrupted or vanish. Blockchain thus seems perversely unsuitable for some of the applications for which it has been proposed. In any situation where one is keeping shared and publicly-visible records that need to be accurate and stable for long periods of time, a blockchain is a poor choice.

*Bankruptcy principles*

It is tempting to consider a poorly-functioning blockchain system near the end of its useful life as comparable to an insolvent person or organization. We might wish to find some counterpart to bankruptcy that can be invoked by honest players, in much the same way that a debtor can resort to bankruptcy or a creditor can force a debtor into bankruptcy. However the change is caused, bankruptcy represents a new operating regime in which rules are radically changed and an outside authority is supervising some kind of reorganization. Instead of being concerned with repayment of debts and preservation of capital, the focus of a hypothetical "blockchain bankruptcy" would be the accurate preservation of archival information despite the faltering nature of the associated blockchain.

However, with a blockchain system there is typically no technical or legal framework to do such a bankruptcy or reorganization. The decentralized nature of the system makes it impossible to impose supervision. The problems identified in previous sections make it hard for any "final" or "reorganized" version of the blockchain to take the place of the failing live system.

Even if we had such a framework in place, it still seems technically challenging to ensure that good-faith queries about a failed blockchain are successfully redirected to the archive. The nature of a blockchain system suggests that dishonest nodes could successfully spoof or hijack such queries. Given that such dishonest players are already showing they are not inclined to behave correctly, additional legal framework is unlikely to improve the situation. Instead, we may need to think in terms of purely technical means to ensure the integrity of a blockchain that is no longer active.

**Related Work**

Potential node misbehavior is a fundamental consideration in essentially all blockchain systems. However, such concerns typically focus on preventing attacks during ordinary system operation. There is considerably less attention paid to issues of thin communities and even less attention to the question of winding down a system.

However, some authors [CKW+16, TE18, A19] have noted unexpected effects or incentives for nodes in a likely future configuration of Bitcoin. When block "mining" rewards are no longer dominant, nodes will earn rewards primarily by transaction fees. Nodes will maximize that reward by changing their behavior: instead of constantly working to win the next block reward, they may wait until a sufficient number of pending transactions (with sufficient associated fees) has accumulated. Even though the potentially-interested community is thick, the active community at a particular point in time may be very thin because everyone is waiting for more pending transactions to arrive. Our observation is that these thin-community problems (caused in these cases by the design of Bitcoin) are all but inevitable near the end of the working life of some blockchain systems.

The phenomenon of "altcoin infanticide" [BMC+15] is a concrete demonstration of the hazards of a thin community. At the startup of a new cryptocurrency, participants in the community of a rival/competitive cryptocurrency may be motivated to eliminate the new competitor. They can potentially do so by dominating the still-thin community of the new cryptocurrency and ensuring that it fails. This phenomenon also shows that it may be too generous to think that a diminishing cryptocurrency is not as much of a problem as non-currency blockchain systems. If coin infanticide is a problem, then it seems plausible that coin geronticide is likewise a potential problem.

Although most successful attacks on blockchain systems are actually attacks on badly-constructed exchanges or buggy smart contracts, there is at least one case where a "real" blockchain (Ethereum Classic, EC) appears to have suffered a 51% attack. [B19] Part of the problem there was that EC had forked from Ethereum and had substantially fewer miners – in our terminology, the EC community became thinner.

**Conclusion**

The difficulty of solving the shutdown problem suggests that blockchains are not a good match for storing data of long-term significance. In particular, storing such data on a blockchain that uses unstable consensus – like Bitcoin – is unwise.

New systems using a blockchain to store data of long-term significance need to be designed with attention to solving the shutdown problem. Existing systems may well need examination to understand their risks, possibly followed by hard-fork upgrades to use more stable consensus.